\documentclass [11pt,a4paper] {article}

\usepackage[cp1252]{inputenc}
\usepackage[english]{babel}

\usepackage{amssymb}
\usepackage{amsmath}
\usepackage{amsfonts,amssymb}

\usepackage[dvips]{graphicx}

\begin{document}

\title{The Computational Limit to Quantum Determinism\\ and the Black Hole Information Loss Paradox}

\author{Arkady Bolotin\footnote{$Email: arkadyv@bgu.ac.il$} \\ \textit{Ben-Gurion University of the Negev, Beersheba (Israel)}}

\maketitle

\begin{abstract}\noindent The present paper scrutinizes the principle of quantum determinism, which maintains that the complete information about the initial quantum state of a physical system should determine the system's quantum state at any other time. As it shown in the paper, assuming the strong exponential time hypothesis, SETH, which conjectures that known algorithms for solving computational NP-complete problems (often brute-force algorithms) are optimal, the quantum deterministic principle cannot be used generally, i.e., for randomly selected physical systems, particularly macroscopic systems. In other words, even if the initial quantum state of an arbitrary system were precisely known, as long as SETH is true it might be impossible in the real world to predict the system's exact final quantum state. The paper suggests that the breakdown of quantum determinism in a process, in which a black hole forms and then completely evaporates, might actually be physical evidence supporting SETH.\\

\noindent \textbf{Keywords:} Determinism, Schrödinger's equation, Computational complexity, NP-complete problems, Exact algorithms, Strong exponential time hypothesis, Information loss paradox.\\
\end{abstract}

\section{Introduction}

\noindent According to the deterministic principle, complete information about a physical system at one point in time should determine its state at any other time. Since all physical systems evolve in time according to the Schrödinger equation $i\hbar\:{\partial\!\!\left.\left|\Psi \!\left(t\right)\!\right.\right\rangle}/{\partial t}=H\!\left(t\right)\!\!\left.\left|\Psi \!\left(t\right)\!\right.\right\rangle$, where $\left.\left|\Psi \!\left(t\right)\!\right.\right\rangle$ is the time-dependent state vector of a system and $H\!\left(t\right)$ is the system's time-dependent Hamiltonian, this means that one can in principle solve this equation for the given physical system with the initial condition $\left.\left|\Psi \!\left(0\right)\!\right.\right\rangle$ to predict the state of the system $\left.\left|\Psi \!\left(t\right)\!\right.\right\rangle$ at any future time $t$.\\

\noindent If we insist that not only a deterministic, unitary evolution but also a wavefunction collapse should be explained due to the Schrödinger equation, then the future state of the system $\left.\left|\Psi \!\left(t\right)\!\right.\right\rangle$ would always be uniquely determined through the linear map\\

\begin{equation} \label{1} 
   \begin{array}{cl}
      \forall H\!\left(t\right) \ \mathcal T:
      &
      \left.\left|\Psi \!\left(t\right)\!\right.\right\rangle \leftarrow \left.\left|\Psi \!\left(0\right)\!\right.\right\rangle
      \;\;\;\; 
   \end{array}  
\end{equation}
\smallskip

\noindent defined by the effect of the time evolution operator $U\!\left(H\!\left(t\right)\!,t,0\right)$ on the initial state of the system $\left.\left|\Psi \!\left(0\right)\!\right.\right\rangle$:\\

\begin{equation} \label{2} 
   \mathcal T \!\left(\left.\left|\Psi \!\left(0\right)\!\right.\right\rangle \!\right) := U\!\left(H\!\left(t\right)\!,t,0\right)\!\left.\left|\Psi \!\left(0\right)\!\right.\right\rangle
\;\;\;\;  ,
\end{equation}
\smallskip

\noindent where the evolution operator $U\!\left(H\!\left(t\right)\!,t,0\right)$ can, in the most general case, be written as\\

\begin{equation} \label{3} 
    U\!\left(H\!\left(t\right)\!,t,0\right)
   =
    \hat \omega \exp  \left(
                                               -\frac{i}{\hbar}\int^t_0\!\!H\!\left(\tau\right)d\tau
                                    \right)
\;\;\;\;  ,
\end{equation}
\smallskip

\noindent provided that $\hat \omega$ is the time-ordered operator. Even though the Schrödinger equation cannot predict the exact result of each measurement but only the probability of these results, the linear mapping (\ref{1}) represents the strictest form of determinism known in physics since it gives all the information about the system for any particular moment of time.\\

\noindent However, the drawback of the mapping (\ref{1}) is that it completely ignores the amount of time (or the number of elementary operations) required to \textit {actually solve} the Schrödinger equation for the given system.\\

\noindent To make this point clearer, let us consider the following scenario: An experimenter conducts an experiment involving an observation of a physical system at some point in time while a theoretician does the parallel calculation using the Schrödinger equation for the given system. At the initial point in time $t=0$ the experimenter sets up the apparatus as the theoretician sets up the system's initial state vector $\left.\left|\Psi \!\left(0\right)\!\right.\right\rangle$. Then, while the experimenter turns on the apparatus and monitors its functioning, the theoretician computes the evolution of the state vector $\left.\left|\Psi \!\left(t\right)\!\right.\right\rangle$ for the system according to the Schrödinger equation. It is clear that in order to predict the result of the observation at the moment $t$, the theoretician must finish up the calculation of the vector $\left.\left|\Psi \!\left(t\right)\!\right.\right\rangle$ ahead of that moment $t$ (i.e., before the experimenter sings out that the observation has occurred and the output is ready).\\

\noindent It is naturally to assume that the vector $\left.\left|\Psi \!\left(t\right)\!\right.\right\rangle$ has an algorithm, i.e., that Schrödinger's equation is solvable. Note that an algorithm here is understood in the sense of the Church–Turing thesis, that is, as a sequence of steps the theoretician with unlimited time and an infinite supply of pen and paper could follow.\\

\noindent Let $\mathcal{A}$ denote an exact algorithm for calculating the effect of the time evolution operator $U\!\left(H\!\left(t\right)\!,t,0\right)$ on the given initial state $\left.\left|\Psi \!\left(0\right)\!\right.\right\rangle$ of the system characterized by the Hamiltonian $H\!\left(t\right)$. Suppose the amount of time taken by this algorithm is not greater than $T$.\\

\noindent Then, according to the deterministic principle (applicable to all physical systems), at any moment $t>0$ the state of every physical system can be determined by the linear map\\

\begin{equation} \label{4} 
   \begin{array}{cl}
      \forall H\!\left(t\right) \ \mathcal T:
      &
      \left.\left|\Psi \!\left(t\right)\!\right.\right\rangle \stackrel{\;\; T}{\leftarrow} \left.\left|\Psi \!\left(0\right)\!\right.\right\rangle
      \;\;\;\;  ,
   \end{array}
\end{equation}
\smallskip

\noindent which explicitly indicates that to associate the state vector $\left.\left|\Psi \!\left(t\right)\!\right.\right\rangle$ of the given system with its initial state $\left.\left|\Psi \!\left(0\right)\!\right.\right\rangle$ the algorithm  $\mathcal{A}$ takes maximally (i.e., in the worst case) the amount of time $T$.\\

\noindent Understandably, the upper bound $T$ may in general depend on the number of the system's constituent microscopic particles $N$, and therefore it can be posed as a function $T\!\left(\!N\right)$, whose behavior is determined by the worst-case complexity of a given Schrödinger's equation (to be exact, by the worst-case complexity of a specific Schrödinger Hamiltonian). Thus the mapping  (\ref{4}) can be rewritten so as to openly contain the number $N$\\

\begin{equation} \label{5} 
   \begin{array}{cl}
      \forall H\!\left(N,t\right) \ \mathcal T:
      &
      \left.\left|\Psi \!\left(t\right)\!\right.\right\rangle \stackrel{\;T\left(\!N\!\right)}{\longleftarrow\!\!\!-} \left.\left|\Psi \!\left(0\right)\!\right.\right\rangle
      \;\;\;\;  .
   \end{array}
\end{equation}
\smallskip

\noindent This modified form of quantum determinism is clearly more realistic than that depicted in (\ref{1}) since it allows for the limit on computational speed in the physical world.\\

\noindent In fact, the form (\ref{1}) implies that there is an algorithm, which can solve Schrödinger's equation either instantaneously or so fast that algorithm's running time $T(N)$ can be ignored\\

\begin{equation} \label{6} 
   \begin{array}{cl}
      \forall H\!\left(N,t\right) \ \mathcal T:
      &
      \left.\left|\Psi \!\left(t\right)\!\right.\right\rangle \stackrel{\;T\left(\!N\!\right)=0}{\longleftarrow\!\!\!-\!\!\!-\!\!\!-\!\!\!-\!\!\!-\!\!\!-} \left.\left|\Psi \!\left(0\right)\!\right.\right\rangle
   \end{array}
\;\;\;\;   
\end{equation}
\smallskip

\noindent Undeniably, in the real world the worst-case running time $T\!\left(\!N\!\right)$ can never be equal to zero, and so $T\!\left(\!N\!\right)>0$.\\

\noindent On the other hand, the deterministic principle demands that the worst-case running time $T\!\left(\!N\!\right)$ can never be greater than the time of observation $t$ – otherwise using the vector $\left.\left|\Psi \!\left(t\right)\!\right.\right\rangle$ to predict the state of the system at the moment $t$ would make no sense. Moreover, the algorithm $\mathcal  A$, which the theoretician uses for solving exactly Schrödinger's equation, would be similarly useless for the purpose of prediction even if the algorithm's worst-case running time $T\!\left(\!N\!\right)$ were equal to the time of observation $t$.\\

\noindent It follows then that for any given physical system the quantum deterministic principle will be valid in the real world only if the running time $T\!\left(\!N\!\right)$ of the algorithm $\mathcal  A$ meets the condition\\

\begin{equation} \label{condition} 
   0<T\!\left(\!N\!\right)<t
\;\;\;\;  .
\end{equation}
\smallskip

\noindent So, the question naturally arises: Can the quantum deterministic principle (\ref{5}) be achievable for all physical systems? In other words, what is the limit, if any, to quantum determinism?\\

\noindent The answer to this question may play the crucial role in dissolving the black hole information loss paradox. This paradox results from the breakdown of unitarity implied by information loss within a black hole.\\

\noindent Imagine a macroscopic system in a pure quantum state that is thrown into a black hole. According to Hawking, the black hole evaporates due to thermal radiation \cite{Hawking75,Hawking76}. Suppose that the black hole continues to evaporate until it disappears completely. As the detailed form of Hawking's radiation does not depend on the detailed structure of the macroscopic system that collapsed into it, we just found a process that converts a pure state into a mixed state \cite{Preskill, Almheiri, Braunstein, Ashtekar}. However, it is clear that transforming a pure quantum state into a mixed state, one must throw away information. Thus, as it turns out the black hole apparently performs a non-unitary transformation on the state of the falling macroscopic system \cite{Okon, Mathur, Gambini1}.\\

\noindent As it is understood now, such a paradox is to a large extent independent from a quantum treatment of the space–time degrees of freedom, i.e. a quantum theory of gravity, but depends crucially on assuming a limitless feasibility of the quantum deterministic principle \cite{Gambini1, Maldacena}. Indeed, if one were willing to drop unitarity then information loss would be no longer problematic \cite{Crull}. Therefore, by demonstrating that quantum determinism cannot be realizable for macroscopic systems, the black hole information loss paradox might be resolved.\\

\noindent The present paper will attack the principle of quantum determinism to demonstrate that this principle formulated in the form of the linear mapping (\ref{5}) is incapable of being used generally, i.e., for randomly selected physical systems, especially macroscopic systems.\\

 \section{Applying the quantum deterministic principle to an adiabatic system}

\noindent Suppose that in the experiment conducted by the experimenter and theoretician the observed physical system $\mathcal M$ evolves slowly from the known prepared ground state $\left.\left|\Psi \!\left(t_{\mathrm{init}}\right)\!\right.\right\rangle$ of the initial Hamiltonian $H_{\mathrm{init}}$  to the ground state $\left.\left|\Psi \!\left(t_{\mathrm{final}}\right)\!\right.\right\rangle$ of another Hamiltonian $H_{\mathrm{final}}$  (not commuting with $H_{\mathrm{init}}$ ) that encodes the solution to some computationally hard problem.\\

\noindent Say, this computational problem is NP-complete (such as the 3SAT problem, the traveling salesman problem, or any other “famous” NP-complete problem discussed in \cite{Karp}). This means that all NP problems (i.e., decision problems with only yes-no answers whose “yes” solutions can be verified in polynomial time) are polynomial-time reducible to this problem. Therefore, finding an efficient algorithm for the given NP-complete problem implies that an efficient algorithm can be found for all NP problems, since any problem belonging to the class NP can be recast into any other member of this class (a brief introduction to the classical theory of computational complexity can be found in \cite{Garey, Papadimitriou}).\\

\noindent To the end that the final Hamiltonian $H_{\mathrm{final}}$  may encode a NP-complete problem, the evolution of the system $\mathcal M$ should take place over the parameter $s=\left(t-t_{\mathrm{init}}\right)/\left(t_{\mathrm{final}}-t_{\mathrm{init}}\right)\in\{0,1\}$ as $H(s)=(1-s)H_{\mathrm{init}}+sH_{\mathrm{final}}$, where  $H_{\mathrm{final}}$  is the quantum version of the Hamiltonian function $H\!\left({\sigma}_1,\dots ,{\sigma}_N\!\right)$ describing the energy of configuration of a set of $N$ spins $\sigma\!{_j}\in\left\{\!-1, +1\right\}$ in the classical Ising model \cite{Fischer, Harris}\\

\begin{equation} \label{8} 
   H\!\left({\sigma}_1,\dots ,{\sigma}_N\!\right) 
   =
  -\sum_{j<k}{C_{jk}\sigma\!{_j}\sigma\!{_k}}
  -A\sum^N_{j}{B_{j}\sigma\!{_j}}
  \;\;\;\;
  \left(A,B_{j},C_{jk}=\rm{const} \right)
\;\;\;\; .
\end{equation}
\smallskip

\noindent One of the computational problems associated with (\ref{8}) is to find the ground state energy of the Hamiltonian function $H\!\left({\sigma}_1,\dots ,{\sigma}_N\!\right)$. Such a function problem can be easily turned into the decision problem: Given the particular choice of the constants $A$, $B_{j}$ and $C_{jk}$, does the ground state of $H\!\left({\sigma}_1,\dots ,{\sigma}_N\!\right)$ have zero energy? Because this decision problem is known to be NP-complete \cite{Barahona, Hartmann}, there exists a polynomial time mapping from this problem to any other NP-complete problem. But then the fact that some other decision problem is NP-complete would mean that it is possible to find a mapping from that problem to the decision problem of the Ising model (\ref{8}) with only a polynomial number of spins ${\sigma}_j$ (see for detail the paper \cite{Lucas} demonstrating that in each case, the required number of spins would be at most cubic in the size of the problem). Consequently, any given NP-complete problem can be written down as the Ising model (\ref{8}).\\

\noindent Let the time interval $t_{\mathrm{final}}-t_{\mathrm{init}}$ be long enough to ensure that the probability of finding the system $\mathcal M$ in the ground state of the final Hamiltonian $H(1)=H_{\mathrm{final}}$  at the end of evolution (i.e., at the time $t_{\mathrm{final}}$) would be close to one. Consider the final Hamiltonian $H_{\mathrm{final}}=H\!({\sigma}_1^z,\dots ,{\sigma}_N^z\!)$, in which spins $\sigma\!{_j}$ of the classical Hamiltonian (\ref{8}) have been replaced by Pauli spin-1/2 matrices ${\sigma}_j^z$. If the resultant quantum Hamiltonian has the zero energy ground state $H\!({\sigma}_1^z,\dots ,{\sigma}_N^z\!)\left.\left|\Psi \!\left(t_{\mathrm{final}}\right)\!\right.\right\rangle=0$, it would mean that there is a solution to the NP-complete problem encoded in the particular Ising model (\ref{8}).\\

\noindent Thus, the application of the quantum deterministic principle to the described quantum system $\mathcal M$ demands that the amount of time $T(N)$ taken by the theoretician in order to predict whether a NP-complete problem encoded in $H(1)$ would have a solution must be less than the evolution time $T_{\mathrm{adiabatic}}=t_{\mathrm{final}} -t_{\mathrm{init}}$  of the observed quantum adiabatic algorithm\\

\begin{equation} \label{9} 
   T\!\left(\!N\!\right)<T_{\mathrm{adiabatic}} \;\; \mathrm{as} \; N \rightarrow \infty
\;\;\;\;  .
\end{equation}
\smallskip

\noindent Let us assess whether such a condition can be always fulfilled.\\

\noindent Though the exact running time $T_{\mathrm{adiabatic}}$ of the adiabatic computation is unknown (it depends on the minimum gap $g=E_1 (s)-E_0 (s)$ between the two lowest levels $E_1(s)$ and $E_0(s)$ of the Hamiltonian $H(s)$ and on its scaling with $N$ \cite{Bapst}), there is evidence \cite{vanDam,Reichardt} that the quantum adiabatic algorithm takes exponential time in the worst-case for NP-complete problems.\\

\noindent Therefore, let us assume that the evolution time $T_{\mathrm{adiabatic}}$ coincides with the maximal amount of time required to trivially solve the NP-complete problem encoded in $H(1)$.\\

\noindent Evidently, to assure the fulfillment of the quantum deterministic principle in that case, the algorithm $\mathcal A$, which the theoretician uses for exactly solving Schrödinger's equation (i.e., for finding $\left.\left|\Psi \!\left(t_{\mathrm{final}}\right)\!\right.\right\rangle$), must be faster than the trivial algorithm. Therefore, the question becomes, does there exist an exact algorithm $\mathcal A$ that can solve the given NP-complete problem faster than brute force?\\

\noindent Here the trouble is that the answer to this question remains unknown: While many NP-complete problems admit algorithms that are much faster than trivial ones, for other problems such as $k$-CNF-SAT, $d$-Hitting Set, or the set splitting problem, no algorithms faster than brute force have been discovered yet (see \cite{Patrascu, Fomin, Lokshtanov} for detail information on exact algorithms for NP-complete problems). Such a situation caused to formalize the hypothesis called the Strong Exponential Time Hypothesis, SETH, which conjectures that certain known brute-force algorithms for solving NP-complete problems are already optimal. More specifically, SETH states that for all $\delta<1$ there is a $k$ (the maximum clause length) such that the $k$-CNF-SAT problem cannot be solved in $O\!\left(2^{\delta N}\!\right)$ time \cite{Impagliazzo62, Impagliazzo63, Calabro, Dantsin}.\\

\noindent Despite the fact that there is no universal consensus about accepting SETH (compared, to say, accepting the P$\ne$NP conjecture), SETH has a special consequence for the quantum deterministic principle.\\

\noindent Indeed, suppose the NP-complete problem encoded in $H(1)$ is the 3-CNF-SAT problem (i.e., a satisfiability problem written as a 3SAT problem in conjunctive normal form). If the strong exponential time hypothesis were true, then this problem could not be exactly solved in time less than the trivial algorithm's running time $O(2^N)$. Thus, if $T_{\mathrm{adiabatic}}=O(2^N)$ then it would necessitate that $T(N)\not<T_{\mathrm{adiabatic}}$, meaning that the quantum deterministic principle could not be fulfilled.\\

\noindent As follows, assuming SETH, quantum determinism cannot be a general principle applicable to all conceivable instances of the quantum adiabatic system $\mathcal M$ described above.\\

\section{Macroscopic quantum determinism}

\noindent Let an ordinary macroscopic system (i.e., a system of Newtonian physics – the physics of everyday life) be characterized by the Schrödinger Hamiltonian $H(N_{\!M})$, where $N_{\!M}$ stands for the number of constituent microscopic particles of such a system. It is safe to assume that $N_{\!M}$ has the order of magnitude, at least, the same as Avogadro's number $N_{\!\rm A}\sim {10}^{24}$.\\

\noindent Since the ordinary macroscopic system has an enormous number of microscopic degrees of freedom, the Hamiltonian  $H(N_{\!M})$ should be \textit{complex enough} to be presented as a sum of $S$ non-overlapping and non-empty terms $H_i(N_i)$\\

\begin{equation} \label{10} 
   H(N_{\!M})=\sum^S_{i=1}{H_i(N_i)}=H_{S^{\prime}}+H_{S^{\prime\prime}}
  \;\;\;\;
  \left(N_i \le N_{\!M}\right)
   \;\;\;\;  
\end{equation}
\smallskip

\noindent such that at least some $S^{\prime}\le S$ of those terms $H_i(N_i)$ would be able to encode computational NP-complete problems (similar to the Hamiltonian function (\ref{8}) of the classical Ising model):\\

\begin{equation} \label{11} 
   H_k(N_k) \le^P  X_k(N_k)
  \;\;\;\;
  \left(k \in \{1,\dots , S^{\prime}\}\right)
   \;\;\;\;  ,
\end{equation}
\smallskip

\noindent where the expression (\ref{11}) denotes a polynomial time reduction from a NP-complete problem $X_k(N_k)$ of size $N_k$ to a Hamiltonian term $H_k(N_k)$. This way, predicting the quantum state $\left.\left|\Psi \!\left(t\right)\!\right.\right\rangle$ of the macroscopic system would require solving the set of NP-complete problems $\mathcal X=\{X_k(N_k)\}^{S^{\prime}}_k$ encoded in the Hamiltonian $H_{S^{\prime}}$.\\

\noindent Unlike degrees of freedom of a microsystem, which can be controlled by the experimenter, the microscopic degrees of freedom of a macroscopic system are mostly out of control. This means that the precise identification of microscopic degrees of freedom governing a macroscopic system's evolution would be impossible. One can infer from here that it is impossible to know with certainty what particular problems $X_k(N_k)$ are enclosed in the set $\mathcal X$. Next it follows that only a generic exact algorithm $\mathcal A$ solving any NP-complete problem would be able to guarantee (even if in principle) the prediction of the exact quantum state $\left.\left|\Psi \!\left(t\right)\!\right.\right\rangle$ of a macroscopic system.\\

\noindent But then again, if SETH held true, there would be no generic exact sub-exponential time algorithm capable of solving all NP-complete problems in sub-exponential or quasi-polynomial time. Consequently, in the worst case, when predicting the exact quantum state $\left.\left|\Psi \!\left(t\right)\!\right.\right\rangle$ the algorithm $\mathcal A$ can converge only in an exponential (or perhaps even larger) amount of time $T_j(N_j )$:\\

\begin{equation} \label{12} 
   T_j(N_j ) =  \max{\{T_k(N_k)\}^{S^{\prime}}_k}
  \;\;\;\;
  \left(j \in \{1,\dots , S^{\prime}\}\right)
   \;\;\;\; ,
\end{equation}
\smallskip

\noindent where $N_j$ is likely to have the same scale as $N_{\!M}$.\\

\noindent Thus, assuming SETH, the principle of quantum determinism would be incapable of being implemented for an arbitrary macroscopic system since it is impossible for an exponential (or faster growing function) $T_j(N_j)$ of a value $N_j$, which has a good chance of being of the same size as Avogadro’ number, to meet the condition $ T_j(N_j )<t$  at any reasonable time $t$.\\

\noindent In other words, even if the initial quantum state $\left.\left|\Psi \!\left(0\right)\!\right.\right\rangle$ of the ordinary macroscopic system were precisely known, as long as SETH is true it would be impossible to predict the system's exact final quantum state $\left.\left|\Psi \!\left(t\right)\!\right.\right\rangle$ in the realm of actual experience.\\

\section{The loss of the information about the initial quantum state by a macroscopic system}

\noindent To be sure, even if SETH held true, a trivial (brute force) way of solving Schrödinger's equation might be nonetheless feasible. Besides the obvious case of a system composed of a few constituent particles completely isolated from the environment, this can be true if there exists a system-specific heuristic that can be used to drastically reduce the system's set of all possible candidates for the witness.\\

\noindent Suppose a macroscopic system $\mathcal M$ to be formally divided into a collective system $\mathcal C$ represented by a small set of the system's collective (macroscopic) observables (along with their conjugate partners) correspond to properties of the macroscopic system $\mathcal M$ as a whole and the environment $\mathcal E$, which is the set of the system's observables other than the collective ones.\\

\noindent It was already noted that the microscopic degrees of freedom of an ordinary macroscopic system are uncontrolled for the most part. It means that one cannot hope to keep track of all the degrees of freedom of the environment $\mathcal E$. Such an inference may be used as a heuristic allowing an enormous set of all possible candidate solutions for $\mathcal M$ to be reduced to just a small set comprising only candidate solutions for $\mathcal C$. Upon applying this heuristic by way of “tracing out” the degrees of freedom of the environment $\mathcal E$ and assuming that the environmental quantum states $\left.\left|\epsilon_n \!\left(t\right)\!\right.\right\rangle$ are orthogonal (or rapidly approach orthogonality), that is, $\left.\langle \epsilon_n \!\left(t\right)\!\left|\epsilon_m \!\left(t\right)\!\right.\right\rangle \to\delta_{nm}$, one would get an inexact yet practicable solution to Schrödinger's equation approximately identical to the corresponding mixed-state density matrix of the system  $\mathcal C$ describing the possible outcomes of the macroscopic observables of the system  $\mathcal M$ and their probability distribution.\\

\noindent As it can be readily seen, the above-described heuristic represents a non-unitary transformation of a pure quantum state into a mixed state (i.e., a probabilistic mixture of pure states) that can be written down as the mapping  $\mathcal A$\\

\begin{equation} \label{13} 
   \begin{array}{cl}
      \mathcal A:
      &
      \rho_{\mathcal C}\!\left(t\right)=
                                \sum_n{\!\left.\left|\phi \!\left(t\right)\!\right.\right\rangle \! c_n c^*_n\!\left.\langle \phi \!\left(t\right)\right|}
                                                \stackrel{\;T\left(\!N_{\mathcal C}\!\right)<t}{\longleftarrow\!\!\!-\!\!\!-\!\!\!-\!\!\!-} 
      \left.\left|\phi_{\mathcal C} \!\left(0\right)\!\right.\right\rangle =
                                \sum_n{\! c_n\!\left.\left|\phi_n \!\left(0\right)\!\right.\right\rangle}
      \;\;\;\; .
   \end{array}
\end{equation}
\smallskip

\noindent where the vector $\left.\left|\phi_{\mathcal C} \!\left(0\right)\!\right.\right\rangle$ and the density operator $\rho_{\mathcal C}\!\left(t\right)$ describe the initial state and the final state of the collective system $\mathcal C$, correspondingly; $N_{\mathcal C}$ stands for the cardinality of the set of all possible candidates for the witness of the system $\mathcal C$.\\

\noindent The loss of information depicted in the mapping (\ref{13}) is especially noteworthy since it cannot be regained. Indeed, to recover the information about phase correlation between different terms in the initial superposition $\left.\left|\phi_{\mathcal C} \!\left(0\right)\!\right.\right\rangle =\sum_n{\!\left.\left|\phi_n \!\left(0\right)\!\right.\right\rangle}$  lost from the collective system $\mathcal C$ to the environment $\mathcal E$, one has to compute the exact total quantum state $\left.\left|\Phi \!\left(t\right)\!\right.\right\rangle =\sum_n{\!c_n\!\left.\left|\phi_n \!\left(t\right)\!\right.\right\rangle \!\left.\left|\epsilon_n \!\left(t\right)\!\right.\right\rangle}$, i.e., to exactly solve the Schrödinger equation for the macroscopic system  $\mathcal M$. But unless SETH falls, solving exactly this equation for an arbitrary macroscopic system can be done only in an exponential, as a minimum, amount of time $T(N_{\!M})$. Therefore – in view of the implausible complexity-theoretic consequences, which the fall of SETH would have for several NP-complete problems \cite{Cygan} – it is highly unlikely that for an ordinary macroscopic system the loss of the information about the initial quantum state might be recovered in any reasonable time.\\

\section{Concluding remarks}

\noindent As it follows from the above discussion, the limit to quantum determinism and the strong exponential time hypothesis \textit{stay and fall together}: If SETH holds, then quantum determinism has a limit since it cannot be a general principle feasibly applicable to any physical system (or to any instance of every physical system). Conversely, if quantum determinism were such a general principle, then SETH could not be valid since for each NP-complete problem there would exist an exact algorithm capable of solving this problem faster than brute force.\\

\noindent Along these lines, the breakdown of the quantum deterministic principle in a process, in which a black hole forms and then completely evaporates, can actually be physical evidence that supports the strong exponential time hypothesis (and thus the P$\ne$NP conjecture).\\

\end{document}